\documentstyle[12pt]{article}
\begin{document}
\thispagestyle{empty}
\newcommand{\p}[1]{(\ref{#1})}
\newcommand{\nn}{\nonumber}
\newcommand{\be}{\begin{equation}}
\newcommand{\ee}{\end{equation}}
\newcommand{\sect}[1]{\setcounter{equation}{0}\section{#1}}
\newcommand{\Db}{{\overline D}}
\newcommand{\Fb}{{\overline F}}
\newcommand{\Gb}{{\overline G}}
\newcommand{\bea}{\begin{eqnarray}}
\newcommand{\eea}{\end{eqnarray}}
\newcommand{\NP}[1]{Nucl.\ Phys.\ {\bf #1}}
\newcommand{\PL}[1]{Phys.\ Lett.\ {\bf #1}}
\newcommand{\CMP}[1]{Comm.\ Math.\ Phys.\ {\bf #1}}
\newcommand{\MPL}[1]{Mod.\ Phys.\ Lett.\ {\bf #1}}
\newcommand{\IJMP}[1]{Int.\ Jour.\ of\ Mod.\ Phys.\ {\bf #1}}
\newcommand{\JMP}[1]{J.\ Math.\ Phys.\ {\bf #1}}
\newcommand{\LMP}[1]{Lett.\ in\ Math.\ Phys.\ {\bf #1}}
\renewcommand{\thefootnote}{\fnsymbol{footnote}}
\newpage
\setcounter{page}{0}
\pagestyle{empty}
\begin{flushright}
{January 1998}\\
{IFT UWr 919/98}\\
{solv-int/9802003}
\end{flushright}
\vskip 10pt
\begin{center}
{\LARGE {\bf Lax pairs for $N=2,3$ supersymmetric KdV}}\\[0.6cm]
{\LARGE {\bf equations and their extensions}}\\[1cm]

\vskip 8pt

{\large S. Krivonos$^{{a,1}}$, A. Pashnev$^{a,2}$ and Z. 
Popowicz$^{b,3}$}
{}~\\
\quad \\
{\em {$^{(a)}$Bogoliubov Laboratory of Theoretical Physics, JINR,}}\\
{\em 141980 Dubna, Moscow Region, Russia}~\quad\\
{\em $^{(b)}$Institute of Theoretical Physics, University of 
Wroclaw,}\\
{\em pl. M. Borna 9, 50-205 Wroclaw, Poland} \\

\end{center}
\vskip 8pt

\centerline{ {\bf Abstract}}
We present the Lax operator for the $N=3$ KdV
hierarchy and consider its extensions.
We also construct a new infinite family of $N=2$
supersymmetric hierarchies by exhibiting the corresponding super Lax
operators. The new realization of $N=4$
supersymmetry on the two general $N=2$ superfields, bosonic spin 1 
and fermionic spin 1/2, is discussed. 
\vskip 4pt

\vfill
{\em E-Mail:\\
1) krivonos@thsun1.jinr.dubna.su\\
2) pashnev@thsun1.jinr.dubna.su\\
3) ziemek@ift.uni.wroc.pl }
\newpage

\pagestyle{plain}
\renewcommand{\thefootnote}{\arabic{footnote}}
\setcounter{footnote}{0}

\noindent{\bf 1. Introduction.}
During the last years the description of three different infinite 
families of 
$N=2$ supersymmetric integrable hierarchies 
in terms of super Lax operators has been proposed in 
\cite{bks,dg,bks1,ks1}.
The generalization to the matrix case
has been derived in \cite{bks1,ks1}. All these Lax operators $L$ have 
the
following generic form \cite{bks1}
\be
L_{KP}^{red}=\partial+b+\mu D+\sum_{j=-\infty}^{-1}
  \left( a_j\partial
 -\left[ Da_j\right] \Db+\omega_j D\partial-\frac{1}{2}
 \left[ D\omega_j\right]\left[ D,\Db\right]\right)\partial^{j-1} \; ,
\label{kpred}
\ee
which is the general solution of the  reduction constraints 
$[D,L_{KP}^{red}]=0$ \cite{sor1} for the $N=2$
supersymmetric KP hierarchy. Here $b$ ($\mu$) are chiral 
bosonic(fermionic)
$N=2$ superfields while $a_j$ ($\omega_j$) are
generic bosonic(fermionic) ones\footnote{The
equivalent reduction $D L_{KP}^{red}=L_{KP}^{red}\Db =0$ has been 
proposed in
\cite{pop1,dg}.}. The Lax operator \p{kpred} still contains
an infinite number of fields and its further reductions 
\cite{bks1,ks1}
are characterized by a finite number of fields and describe
three families of $N=2$ supersymmetric hierarchies
with $N=2$ super $W_n$ algebras as their second Hamiltonian structure.

These three families of hierachies include the $N=2$ $a=-2,4$ KdV 
equations \cite{lm} and two KdV equations  with $N=4$ $SU(2)$ 
superconformal algebra as the
second Hamiltonian structure \cite{di,ik,dgi,dg}.
However they  do not describe neither $N=2$ $a=1$ KdV 
equation \cite{lm} nor $N=3$ KdV ones \cite{yung}. The aim of this 
letter is 
to present the Lax operator for $N=3$ KdV equation, which generalize 
the Lax 
operator for $N=2$ $a=1$ KdV ones \cite{pop2}, and based on it, to 
construct 
a new infinite class of further reductions (with a finite numbers of 
fields) 
of the Lax operator \p{kpred} which gives rise to new integrable $N=2$
hierarchies.

{}~

\noindent{\bf 2. Lax representation for $N=3$ super KdV equation.}
$N=3$ supersymmetric KdV equation with $N=3$ super conformal algebra 
as
its second Hamiltonian structure has been proposed in \cite{yung}. In
terms of $N=2$ superfields it can be written as the following coupled
system of evolution equations for the general bosonic spin 1 
superfield 
$J(Z)$ and general fermionic spin 1/2 superfield $g(Z)$:
\bea
\frac{\partial}{\partial t} J & = &
-J'''-(J^3)'+3(J\;[D,\overline{D}]J)'
+3(g'\;[D,\overline{D}]g)'- \nn \\
&&12(J \;Dg \;\overline{D}g)'+
6( \overline{D}J\; g\; Dg)'+6(DJ\; g\; \overline{D}g)' \; ,
  \nn \\
\frac{\partial}{\partial t} g & = & 
-g'''+6J(\overline{D}J\; Dg+DJ\; \overline{D}g)+
3(J\;[D,\overline{D}]g)'-3J^2\;g'- \nn \\
&&6g'\; Dg\; \overline{D}g-6(g\; Dg\; \overline{D}g)'\; ,
 \label{n3kdv}  
\eea
where $Z=(z,\theta,\overline\theta)$ is a coordinate of
the $N=2$ superspace, $dZ \equiv dz d \theta d \overline\theta$ and
the fermionic covariant derivatives $D$ and ${\overline D}$ are 
defined as
\begin{eqnarray}
D=\frac{\partial}{\partial\theta}
 -\frac{1}{2}\overline\theta\frac{\partial}{\partial z}, \quad
{\overline D}=\frac{\partial}{\partial\overline\theta}
 -\frac{1}{2}\theta\frac{\partial}{\partial z}, \quad
D^{2}={\overline D}^{2}=0, \quad
\left\{ D,{\overline D} \right\}= -\frac{\partial}{\partial z}
\equiv -{\partial}.
\label{DD}
\end{eqnarray}
One can check that the equations \p{n3kdv} are covariant with respect
to the following transformations of an extra hidden supersymmetry:
\be
\delta J = \varepsilon g', \quad \delta g=\varepsilon J \;,
\ee
where $\varepsilon$ is a Grassmann parameter. Just this additional
supersymmetry together with explicit $N=2$ ones form $N=3$ 
supersymmetry.

The $N=3$ KdV equation \p{n3kdv} can be obtained from the Hamiltonian
\cite{yung}
\bea 
H&=&-3\int\! dZ\left(J[D,\Db ]J+\frac{1}{3}J^3-g[D,\Db ]g'+
 2Jgg'+8JDg\Db g\right) \;, \label{hamn3kdv} \\
\frac{\partial}{\partial t} J & \equiv & \left\{ H,J\right\} , \;
\frac{\partial}{\partial t} g  \equiv  \left\{ H,g\right\} , \nn
\eea
if we use the $N=3$ superconformal algebra Poisson brackets, which
read in terms of $N=2$ supercurrents $J(Z)$ and $g(Z)$ as follows:
\bea
\left\{ J(Z_1),J(Z_2) \right\} & = & 
\left( \frac{1}{2}[D,\overline{D}]\partial+\partial J+J\partial+
\overline{D}JD
+DJ\overline{D}\right) \Delta (Z_1-Z_2), \nn \\
\left\{ J(Z_1),g(Z_2) \right\} & = &  
\left(\partial g
+\frac{1}{2}g\partial-\overline{D}gD-Dg\overline{D}
\right)\Delta(Z_1-Z_2),\nn \\
\left\{ g(Z_1),g(Z_2) \right\} & = & 
\frac{1}{2}\left( J\partial+[D,\overline{D}]\right)\Delta(Z_1-Z_2)
 \;. \label{n3sca}
\eea
Here $\Delta(Z_1-Z_2)=(\theta_1-\theta_2)(\overline{\theta}_1-
\overline{\theta}_2)\delta(z_1-z_2)$ is $N=2$ superspace delta 
function and 
all operators in r.h.s. are evaluated in the second point. 
Let us stress that in the contrast with two other different
$N=3$ KdV equations \cite{cl,bik}, the equation \p{n3kdv} and 
Hamiltonian
\p{hamn3kdv} explicitly break the $SO(3)$ internal symmetry of $N=3$ 
superconformal algebra to $U(1)$ symmetry.

It was shown in \cite{yung} that the equation \p{n3kdv} possesses 
first 3 nontrivial conservation laws and therefore it should be 
integrable 
one. Now we turn to the basic item of this section, the construction 
of the
Lax operator for the $N=3$ KdV hierarchy \p{n3kdv}. Keeping in mind 
that 
in the  limit $g\rightarrow 0$ the $N=3$ KdV equation reduces to the
$N=2$ $a=1$ KdV one, which can be described by the following Lax
operator \cite{pop2}
\be\label{laxa1}
L_{a=1} = \partial - \left[ D, \Db \right] \partial^{-1} J
\ee
we propose the following Lax operator for $N=3$ KdV hierarchy
\be\label{laxn3}
L = \partial - \left[ D, \Db \right] \partial^{-1} J-
\left[ D, \Db \right] \partial^{-1} g
\left[ D, \Db \right] \partial^{-1} g \; .
\ee
Now, one can check that the Lax operator \p{laxn3} indeed gives rise
to the $N=3$ KdV flows via the Lax equation
\be
\frac{\partial}{\partial t_{2n+1}} L =
\left[ \left( L^{2n+1}\right)_{+} ,L\right] \; ,
\ee
where $\{+\}$ denotes the differential part of the Lax operator.
The conserved currents for the Lax operator \p{laxn3} are defined by
the standard $N=2$ residue form as
\be
H_n= \int\! dZ \;tr\left( L^{2n+1}\right) \; ,
\ee
where $tr$ denotes the coefficient standing before $[D,\Db]\partial^{-
1}$.

Thus, we have proved the integrability of the $N=3$  KdV equation.

To close this section, let us note that the field
$g$ appears in the Lax operator \p{laxn3} only in pair. Therefore, 
one can
immediately generalize the Lax operator \p{laxn3} as follows
\be\label{laxn3g}
L = \partial - \left[ D, \Db \right] \partial^{-1} J-
\sum_{i=1}^M \left[ D, \Db \right] \partial^{-1} g_{i}
\left[ D, \Db \right] \partial^{-1} g_{i}-
\sum_{j=1}^K \left[ D, \Db \right] \partial^{-1} b_{j}
\left[ D, \Db \right] \partial^{-1} b_{j}
\ee
where we introduced $M$ general $N=2$ fermionic superfields $g_i$ and
$K$ general $N=2$ bosonic superfields $b_j$. The resulting coupled 
system
of evolution equations look a bit complicated and we write here the 
third 
flow equations only for the case $b_j=0$
\bea
\frac{\partial}{\partial t} J & = & 
-J'''-(J^3)'+3(J\;[D,\overline{D}]J)'-
12(J\; Dg_i\; \overline{D}g_i)'+\nonumber\\
&&6( \overline{D}J\; g_l\; Dg_l)'+6(DJ\; g_l\; \overline{D}g_l)'
-3([D,\overline{D}]g_l\; g_{l}')' \; ,\nn \\
\frac{\partial}{\partial t} g_i & = & 
-g_{i}'''+6J(\overline{D}J\; Dg_i+DJ\; \overline{D}g_i)+
3(J\;[D,\overline{D}]g_i)'-3J^2\;g_{i}'-\nonumber\\
&& 6(g_l\;\overline{D}g_i)'\; Dg_l-6(g_l\; Dg_i)'\;\overline{D}g_l+
3[D,\overline{D}]g_i\; [D,\overline{D}]g_l\; g_l-3g_{i}'\; g_{l}'\; 
g_l
 \label{n3kdvg}  
\eea
(summation over repeated indices is understood).
Thus we have got new integrable extensions of $N=3$ KdV hierarchy.

{}~

\noindent{\bf 3. New Lax representation for $N=4$ KdV.}
In some cases the known Lax operators could give hints how to
construct their generalization or ever how to construct new Lax 
operators.
For example, the $N=4$ KdV Lax operator \cite{ik} has been 
constructed as
junction of two known Lax operators. As an another example, the
$N=2$ $a=4$ KdV lax operator \cite{lm} can be constructed from the 
$N=2$ $a=1$ one \p{laxa1} as follows \cite{pop3}:
\be
L_{a=4}=L_{a=1}-L_{a=1}^{*} \; ,
\ee
where star means formal operator conjugation. Since the $N=3$ KdV
Lax operator \p{laxn3} is a generalization of the $N=2$ $a=1$ KdV one,
the analogous procedure in this case might yield new integrable
systems. The basic aim of this section is to demonstrate that this is
indeed the case.

Thus, based on the above consideration, let us present the following
Lax operator
\begin{eqnarray}
L&=& \partial-J-\overline{D}\partial^{-1}\left[DJ\right]+
 \partial^{-1}\left[\overline{D}Dg\right]
\overline{D}\partial^{-1}\left[Dg\right]+ 
 \partial^{-1}\overline{D}g\left[Dg\right]+
g\overline{D}\partial^{-1}\left[Dg\right]  \; .\label{laxn4}
\end{eqnarray}
Here $J$ and $g$ are the general $N=2$ bosonic and fermionic 
superfields
correspondingly, and the square brackets mean that the relevant 
operators
act only on the superfields inside the brackets. We have checked that
the Lax operator \p{laxn4} gives rise, through the Lax equation
\be
\frac{\partial}{\partial t_n} L =
\left[ \left( L^n\right)_{\geq 1} ,L\right]
\ee
to the self-consistent hierarchy of the evolution equations 
(here subscript $\{\geq 1\}$ denotes the purely differential part of 
the Lax 
operator). 
For the Lax operator \p{laxn4} the Hamiltonians $H_n$ are obtained 
from the constant term of $L^n$ \cite{kst}, that is
\be
H_n= \int\! dZ\; (L^n)_0 \;,
\ee
where subscripts  $0$ means the constant part of an operator.

Explicitly, the second flow reads:
\bea
\frac{\partial}{\partial t_2} J &=&
  [D,\overline{D}]J'+2J'J+2(Dg\; \overline{D}g)'\; , \nn \\
\frac{\partial}{\partial t_2} g & = & [D,\overline{D}]g'-
   2\overline{D}J\; Dg -2DJ\; \overline{D}g+2g'\; J \; . 
\label{n4kdv1}
\eea
An interesting peculiarity of the Lax operator \p{laxn4} is that it
can be rewritten in terms of $J$ and bosonic chiral-antichiral 
superfields
$G,{\overline G}$, defined as
\be
G\equiv Dg,\quad {\overline G}\equiv \Db g \; .
\ee
In this new basis the Lax operator \p{laxn4} reads as
\be
L= \partial -J-\overline{D}\partial^{-1}(DJ)+ 
 \partial^{-1}G\overline{G}- \partial^{-1}(D\overline{G})\partial^{-1}
  \overline{D}G \label{laxn41} \;,
\ee
while the second flow equations take the form
\begin{eqnarray}
\frac{\partial}{\partial t_2} J&=&[D,\overline{D}]J'+
   2(G\;\overline{G})'+2J'J \; ,\nn \\
\frac{\partial}{\partial t_2} G&=&G''-2D\Db(J\;G)\;, \quad
\frac{\partial}{\partial t_2} \overline{G}=-\overline{G}''-
  2\Db D(J\; \overline{G}) \; . \label{n4kdv2}
\end{eqnarray}
In the equations \p{n4kdv2} one can immediately recognize the second 
flow
equations of the $N=4$ KdV hierarchy \cite{di,dik}. Thus, the Lax 
operators
\p{laxn4},\p{laxn41} provide the new description of the $N=4$ KdV 
hierarchy.

It is easy to check the covariance of the second flow 
equations \p{n4kdv2} under the transformations of an extra hidden 
$N=2$
supersymmetry \cite{dik}
\be
\delta J=\epsilon D\overline{G}+
\overline{\epsilon}\overline{D}G\;,\quad
\delta G=\epsilon DJ \;,\quad 
\delta \Gb =\overline{\epsilon}\overline{D}J\;,
\ee
where $\epsilon,\bar\epsilon$ are mutually conjugated Grassmann 
parameters.
Together with explicit $N=2$ supersymmetry they form $N=4$ 
supersymmetry.
It is somewhat surprising that the $N=4$ supersymmetry can be 
realized on the
superfields $J$ and $g$, but in a non-local way
\be
\delta J=\overline{\epsilon}\overline{D}Dg+
\epsilon D\overline{D}g \;,\quad
\delta g=\overline{\epsilon}D\overline{D}\partial^{-1} J+
            \epsilon\overline{D}D\partial^{-1} J \;. 
\ee
Thus the pair of general superfields $J,g$ also forms the multiplet of
$N=4$ supersymmetry. Let us stress ones more that all flow 
equations of $N=4$ KdV hierarchy together with all Hamiltonians can be
rewritten in terms of general superfields $J$ and $g$. We suspect the 
$N=4$ $SU(2)$ superconformal algebra itself can be rewritten in terms 
of
superfields $J$ and $g$.

{}~

\noindent{\bf 4. New reduction of $N=2$ KP hierarchy.} In the 
previous 
section we constructed the new Lax operator for the $N=4$ KdV 
hierarchy. It turns out that this Lax operator \p{laxn4} commutes
with spinor covariant derivative $D$ and then it belongs to the
same class of Lax operators as \p{kpred}. Moreover, it looks like
$N=2$ $a=4$ KdV Lax operator plus extra pieces including
only additional field $g$ (or $G$ and $\Gb$). Therefore, it is natural
to consider the following extension of this Lax operator \cite{bks1}:
\begin{eqnarray}
L_s&=&\partial^s+\sum_{j=1}^{s-1}\left( 
  J_{s-j}\partial-\left[DJ\right]\overline{D}\right)\partial^{j-1} 
 -J_s-\overline{D}\partial^{-1}\left[DU\right]-
  F_A\Fb_A-F_A\Db\partial^{-1}\left[D\Fb_A\right] \nn \\
&&
 +\partial^{-1}\left[\overline{D}Dg_B\right]
 \overline{D}\partial^{-1}\left[Dg_B\right]+ (-1)^{d_{g_{B} }}
 \partial^{-1}\overline{D}g_B\left[Dg_B\right]+
 g_B\overline{D}\partial^{-1}\left[Dg_B\right]  \label{newlax}
 \end{eqnarray}
(summation over repeated indices is understood).
Here $d_{g_B}$ is a grassmann parity of superfield $g_B$, 
$d_g=1\;(d_g=0)$
for fermionic (bosonic) superfields, $J_s$ are general bosonic
$N=2$ superfields,
$F_A$ and $\Fb_A$ are $(n+m)$
pairs of chiral and antichiral $N=2$ superfields
\be
DF_A=\Db\Fb_A=0
\ee
which are fermionic for $A=1,\ldots,n$ and bosonic for 
$A=n+1,\ldots,n+m$
and $g_B$ are $k+l$ pairs of general $N=2$ superfields,
fermionic for $B=1,\ldots,k$ and bosonic for $B=k+1,\ldots,k+l$.
Such operator provides the consistent flows
\be
\frac{\partial}{\partial t_n} L =
\left[ \left( L^n\right)_{\geq 1} ,L\right] \; ,
\ee
and the infinite number of Hamiltonians can be obtained in a standard 
way:
\be
H_n= \int dZ (L^n)_0
\ee

To better understand what kind of hierarchies we have proposed,
let us consider explicitly the first simplest hierarchies 
corresponding to 
the values $s=1,2$ and $s=3$ in the Lax operator $L_s$ \p{newlax}, 
with single fermionic superfield 
$g$ $(k=1,l=0)$ and one pair of $F,\Fb$ superfields $(n=1,m=0)$.

{}~

\noindent{1. The $s=1$ case.}

For this simplest case the Lax operator \p{newlax} has the following 
form
($J_1\equiv J$):
\begin{eqnarray}
L_1&=& \partial -J-\overline{D}\partial^{-1}\left[DJ\right]- 
   F\Fb-F\Db\partial^{-1}\left[D\Fb\right] \nn \\
&& + \partial^{-1}\left[\overline{D}Dg\right]
 \overline{D}\partial^{-1}\left[Dg\right]+ 
 \partial^{-1}\overline{D}g\left[Dg\right]+
 g\overline{D}\partial^{-1}\left[Dg\right] \; ,  \label{ex1}
\end{eqnarray}
while the second flow equations read
\bea
\frac{\partial}{\partial t_2} J &=&
  [D,\overline{D}]J'+2J'J+2(Dg \;\overline{D}g)' 
-2\overline{D}J\; F\; D\overline{F}-2\overline{F}\;F\; J'+
2DJ\;\overline{F}\;\overline{D}F,\nonumber\\
\frac{\partial}{\partial t_2} g & = & [D,\overline{D}]g'-
   2\overline{D}J Dg -2DJ \overline{D}g-2g' J  -
2g'\overline{F}F-2\overline{F}\;Dg\; \overline{D}F+
2F\;\overline{D}g\; D\overline{F}, \nn \\
\frac{\partial}{\partial t_2} F & = & -F''+2F'J+2\overline{F}F'F
+2F\;D\overline{F}\;\overline{D}F-2DJ\;\overline{D}F, \nn \\
\frac{\partial}{\partial t_2} \Fb & = & 
\overline{F}{}''+2\overline{F}{}'J
-2\overline{F}\;D\overline{F}\;\overline{D}F -
      2\overline{F}{}'\;\overline{F}\;F
      -2\overline{D}J\; DF     \label{ex1e}
\eea
From these expressions we can easily see that in the both
limits $F=\Fb=0$ and $g=0$ they coincide with the corresponding flows 
 
of the $N=4$ KdV hierarchy \cite{di,dik} in the different bases while
at $F=\Fb=g=0$ we obtain the $N=2$ $a=4$ KdV hierarchy.
Thus, our family of $N=2$ hierarchies
includes the well-known $N=2$ $a=4$ and $N=4$ KdV hierarchies 
(in two different bases) and possesses the new Lax-pair 
representation for 
their extensions.

{}~

\noindent{2. The $s=2$ case.}

For this case the Lax operator reads as ($J_1\equiv J,J_2\equiv W)$
\begin{eqnarray}
L&=&\partial^2+J\partial -\left[ DJ\right]\overline{D} -W
-\overline{D}\partial^{-1}\left[ DW\right] -
F\overline{F}-F\overline{D}\partial^{-1}
 \left[ D\overline{F}\right]\nonumber\\
&&+\partial^{-1}\left[ \overline{D}Dg\right]\overline{D}
     \partial^{-1}\left[ Dg\right]+
\partial^{-1}\overline{D} g \left[ Dg\right]+
  g\overline{D}\partial^{-1}\left[ Dg\right] \;,
\end{eqnarray}
providing the following second flows equations 
\bea
\frac{\partial}{\partial t_2} J&=&2W'-2(\overline{F}F)'\;,\;
\frac{\partial}{\partial t_2} W=
 [D,\overline{D}]W' -\overline{D}W DJ -DW \overline{D}J+
 2(Dg \overline{D}g)'-W' J \; , \nn \\
\frac{\partial}{\partial t_2} F&=&-F''-F' J+DJ \overline{D}F,\;
\frac{\partial}{\partial t_2} \overline{F}=
 \overline{F}{}''-\overline{F}{}' J +
 \overline{D}J D\overline{F} \; , \nn \\
\frac{\partial}{\partial t_2} g&=&[D,\overline{D}]g' +
\overline{D}J Dg +DJ \overline{D}g-g' J \;.
\eea
With $g=0$ this Lax operator has been considered in \cite{pop3}. With 
the
superfield $g$ it describes new extension of the $N=2$ $a=-2$ super 
Boussinesq hierarchy ($F=\Fb=g=0$ limit).

{}~

\noindent{3. The $s=3$ case.}

For this more complicated case we present only the third--flow 
equations
in the limit $F=\Fb=0$ which give the new supersymmetric extension of
the $N=2$ $a=-1/2$ Boussinesq equation (again we redefined the fields 
as
$J_1\equiv J,J_2\equiv W,J_3\equiv U$):
\bea
\frac{\partial}{\partial t_3}J& = &3U' \;, \nn \\
\frac{\partial}{\partial t_3}W&=&\frac{3}{2} U''-
 \frac{3}{2}[D,\overline{D}]U' +\overline{D}U DJ
  +DU \overline{D}J-
  3Dg'\overline{D}g+2U'J-3Dg \overline{D}g'\;, \nn \\
\frac{\partial}{\partial t_3}U_t&=& -U'''-\overline{D}U' 
DJ+\overline{D}W DU-
 \frac{1}{2}\overline{D}J [D,\overline{D}]g Dg+
 \frac{1}{2}\overline{D}J g' Dg+DU' \overline{D}J \nn \\
 &&+\frac{3}{2}([D,\overline{D}]g g)' 
 +DU \overline{D}J'+
 DW \overline{D}U +DJ [D,\overline{D}]g \overline{D}g
  +DJ g' \overline{D}g\nn \\
&& +[D,\overline{D}]U' J
 +J' Dg \overline{D}g+\frac{1}{2}[D,\overline{D}]U J' +\frac{1}{2}U' 
J'
-W U' +2J (Dg \overline{D}g)' \; , \nn \\
\frac{\partial}{\partial t_3}g&=& -g'''-(\overline{D}J Dg)' +
  [D,\overline{D}]g' J +\frac{1}{2}[D,\overline{D}]g J'+
 \overline{D}W Dg+DW \overline{D}g \nn \\
&&+\frac{1}{2}g' J'-g' W+DJ \overline{D}g' \;.
\end{eqnarray}
The extension of this system by the superfields $F$ and $\Fb$
can be straightforwardly derived from the Lax-pair representation
\p{newlax} and we do not present it here.

{}~

\noindent{\bf 5. Conclusion. }
In this letter we present the Lax operator for $N=3$ KdV hierarchy
and thus proved the integrability of the latter.
We also constructed a new infinite family of $N=2$
supersymmetric hierarchies by exhibiting the corresponding super Lax
operators. As a by-product we find the new realization of $N=4$
supersymmetry on the two general superfields -- bosonic spin 1 $J$
and fermionic spin $1/2$ $g$. We suspect that the constructed
Lax operators can be extended to the matrix case, like in 
\cite{bks1,ks1}
(at least for the Lax operators from the sections 4,5). It is rather
interesting problem to extend the present consideration to the case
of third KP reduction hierarchies \cite{ks1}. The detailed analysis
of these problems is under way.

{}~

\noindent{\bf Acknowledgments.}
S.K. and A.P. thank Institute of Theoretical Physics in Wroclaw for
hospitality during the course of this work.
This work was partially supported by the Russian
Foundation for Basic Research, Grant No. 96-02-17634, RFBR-DFG Grant 
No.
96-02-00180, INTAS Grants  93-127 ext, 96-0308 and 96-0538

\end{document}